\RequirePackage{fix-cm}
\documentclass[smallextended]{svjour3}       
\smartqed  
\usepackage{graphicx,amssymb,amsfonts,mathrsfs,latexsym,amsmath,amssymb}
\journalname{***}
\begin{document}

\title{On $(\alpha+u\beta)$-constacyclic codes of length $p^sn$
over $\mathbb{F}_{p^m}+u\mathbb{F}_{p^m}$
} \subtitle{}

\titlerunning{On $(\alpha+u\beta)$-constacyclic codes of length $p^sn$
over $\mathbb{F}_{p^m}+u\mathbb{F}_{p^m}$}

\author{ Yuan Cao $\cdot$ Qingguo Li
}


\institute{Y. Cao (corresponding author),  Q. Li\at
              College of Mathematics and Econometrics, Hunan University, Changsha 410082, China
               \\
              \email{yuan$_{-}$cao@hnu.edu.cn, liqingguo@hnu.edu.cn}                 
}

\date{Received: date / Accepted: date}

\maketitle

\begin{abstract}
Let $\mathbb{F}_{p^m}$ be a finite field of cardinality $p^m$ and $R=\mathbb{F}_{p^m}[u]/\langle u^2\rangle=\mathbb{F}_{p^m}+u\mathbb{F}_{p^m}$ $(u^2=0)$, where $p$ is an odd prime and $m$ is a positive integer.
For any $\alpha,\beta\in \mathbb{F}_{p^m}^{\times}$, the aim of this paper is to represent all distinct $(\alpha+u\beta)$-constacyclic codes
over $R$ of length $p^sn$ and their dual codes, where $s$ is a nonnegative integer and $n$ is a
positive integer satisfying ${\rm gcd}(p,n)=1$. Especially, all distinct
$(2+u)$-constacyclic codes of length $6\cdot 5^t$
over $\mathbb{F}_{3}+u\mathbb{F}_3$ and their dual codes are listed, where $t$ is a positive integer.

\end{abstract}

\section{Introduction}
\label{intro}
  Algebraic coding theory deals with the design of error-correcting and error-detecting codes for the reliable transmission
of information across noisy channel. The class of constacyclic codes plays a very significant role in
the theory of error-correcting codes as they can be efficiently encoded with simple shift
registers. This family of codes is thus interesting for both theoretical and practical reasons.

\par
  Let $\Gamma$ be a commutative finite ring with identity $1\neq 0$, and $\Gamma^{\times}$ be the multiplicative group of invertible elements of
$\Gamma$. For any $a\in
\Gamma$, we denote by $\langle a\rangle_\Gamma$, or $\langle a\rangle$ for
simplicity, the ideal of $\Gamma$ generated by $a$, i.e., $\langle
a\rangle_\Gamma=a\Gamma=\{ab\mid b\in \Gamma\}$. For any ideal $I$ of $\Gamma$, we will identify the
element $a+I$ of the residue class ring $\Gamma/I$ with $a$ (mod $I$) for
any $a\in \Gamma$.

\par
   A \textit{code} over $\Gamma$ of length $N$ is a nonempty subset ${\cal C}$ of $\Gamma^N=\{(a_0,a_1,\ldots$, $a_{N-1})\mid a_j\in\Gamma, \
j=0,1,\ldots,N-1\}$. The code ${\cal C}$
is said to be \textit{linear} if ${\cal C}$ is an $\Gamma$-submodule of $\Gamma^N$. All codes in this paper are assumed to be linear. The ambient space $\Gamma^N$ is equipped with the usual Euclidian inner product, i.e.,
$[a,b]_E=\sum_{j=0}^{N-1}a_jb_j$, where $a=(a_0,a_1,\ldots,a_{N-1}), b=(b_0,b_1,\ldots,b_{N-1})\in \Gamma^N$,
and the \textit{dual code} is defined by ${\cal C}^{\bot_E}=\{a\in \Gamma^N\mid [a,b]_E=0, \forall b\in {\cal C}\}$.
If ${\cal C}^{\bot_E}={\cal C}$, ${\cal C}$ is called a \textit{self-dual code} over $\Gamma$.

\par
   Let $\lambda\in \Gamma^{\times}$.
A linear code
${\cal C}$ over $\Gamma$ of length $N$ is
called a $\lambda$-\textit{constacyclic code}
if $(\lambda c_{N-1},c_0,c_1,\ldots,c_{N-2})\in {\cal C}$ for all
$(c_0,c_1,\ldots,c_{N-1})\in{\cal C}$. Particularly, ${\cal C}$ is
called a \textit{negacyclic code} if $\lambda=-1$, and ${\cal C}$ is
called a  \textit{cyclic code} if $\lambda=1$.
  For any $a=(a_0,a_1,\ldots,a_{N-1})\in \Gamma^N$, let
$a(x)=a_0+a_1x+\ldots+a_{N-1}x^{N-1}\in \Gamma[x]/\langle x^N-\lambda\rangle$. We will identify $a$ with $a(x)$ in
this paper. By [8] Propositions 2.2 and 2.4 we have

\vskip 3mm \noindent
  {\bf Lemma 1.1}  \textit{Let $\lambda\in \Gamma^{\times}$. Then ${\cal C}$ is a  $\lambda$-constacyclic code
of length $N$ over $\Gamma$ if and only if ${\cal C}$ is an ideal of
the residue class ring $\Gamma[x]/\langle x^N-\lambda\rangle$}.

\vskip 3mm \noindent
  {\bf Lemma 1.2}  \textit{The dual code of a $\lambda$-constacyclic code of length $N$ over
$\Gamma$ is a $\lambda^{-1}$-constacyclic code of length $N$ over
$\Gamma$, i.e., an ideal of $\Gamma[x]/\langle
x^N-\lambda^{-1}\rangle$}.

\vskip 3mm \par
  In this paper, let $\mathbb{F}_{p^m}$ be a finite field of cardinality $p^m$, where
$p$ is a prime and $m$ is a positive integer, and denote $\mathbb{F}_{p^m}[u]/\langle u^2\rangle$
by $\mathbb{F}_{p^m}+u\mathbb{F}_{p^m}$ ($u^2=0$). There were a lot of literatures on linear codes, cyclic codes and
constacyclice codes of length $N$ over rings $\mathbb{F}_{p^m}+u\mathbb{F}_{p^m}$ ($u^2=0$) for various prime $p$ and positive integers $m$ and $N$.
For example, [1,2,4,10,12,13 and 15].
The classification of codes plays an important role in studying their structures and encoders.
However, it is a very difficult task in general, and only some codes of special lengths over certain finite
fields or finite chain rings are classified. For example,
all constacyclic codes of length $2^s$ over the Galois extension
rings of $\mathbb{F}_2 + u\mathbb{F}_2$ are classified and their detailed structures are also established in [7]. In [8], Dinh
classified all constacyclic codes of length $p^s$ over $\mathbb{F}_{p^m}+u\mathbb{F}_{p^m}$. Recently, Dinh et al. [9] studied
negacyclic codes of length $2p^s$ over the ring $\mathbb{F}_{p^m}+u\mathbb{F}_{p^m}$. The purpose of this paper is to continue this
line of research.  We determine the algebraic structures of a class of constacyclic codes of arbitrary length over $\mathbb{F}_{p^m}+u\mathbb{F}_{p^m}$.
   In this paper, we will adopt the following notations:

\vskip 3mm \noindent
   {\bf Notation 1.3} Let $\alpha,\beta\in \mathbb{F}_{p^m}^{\times}$, $p$ be an odd prime, and $m,s,n$ be positive integers
satisfying ${\rm gcd}(p,n)=1$. We denote

\vskip 2mm \par
  $\bullet$ $R=\mathbb{F}_{p^m}[u]/\langle u^2\rangle=\mathbb{F}_{p^m}
+u\mathbb{F}_{p^m}$ ($u^2=0$);

\vskip 2mm \par
   $\bullet$ $\mathcal{R}_{\alpha+u\beta}=R[x]/\langle x^{p^sn}-(\alpha+u\beta)\rangle$ and
    $\mathcal{A}=\mathbb{F}_{p^m}[x]/\langle (x^{p^sn}-\alpha)^2\rangle$.

\vskip 3mm\par
    The present paper is organized as follows.
In Section 2, we provide an explicit generator for each $(\alpha+u\beta)$-contacyclic code over $R$ of length $p^sn$ and give a formula to count the number of codewords in each code.
In Section 3, we determine the generator for the dual code of any $(\alpha+u\beta)$-contacyclic code over $R$ of length $p^sn$.
Especially, we list
all $(2+u)$-constacyclic codes of length $6\cdot 5^t$
over $\mathbb{F}_{3}+u\mathbb{F}_3$ and their dual codes for any positive integer $t$ .


\section{Generators for $(\alpha+u\beta)$-constacyclic codes over $R$ of length $p^sn$}
\noindent
In this section, we determine all distinct $(\alpha+u\beta)$-constacyclic codes over $R$ of length $p^sn$, i.e., all distinct ideals
of the ring $\mathcal{R}_{\alpha+u\beta}$.

\par
   First, we establish a relationship between rings $\mathcal{R}_{\alpha+u\beta}$ and $\mathcal{A}$.
Let $a(x)\in \mathcal{A}$. Then $a(x)$ can be uniquely expressed as $a(x)=\sum_{i=0}^{2p^sn-1}b_ix^i\in\mathbb{F}_{p^m}[x]$ where $b_i\in \mathbb{F}_{p^m}$ for
all $0=0,1,\ldots,2p^sn-1$. Dividing $a(x)$ by
$\beta^{-1}(x^{p^sn}-\alpha)$, we obtain a unique pair $(a_0(x),a_1(x))$ of polynomials $a_0(x)$, $a_1(x)\in \mathbb{F}_{p^m}[x]$ such
that
$$a(x)=a_0(x)+\beta^{-1}(x^{p^sn}-\alpha)a_1(x),$$
where $a_k(x)=\sum_{t=0}^{p^sn-1}b_{k,t}x^t$ with $b_{k,t}\in \mathbb{F}_{p^m}$ for all $t=0,1,\ldots,p^sn-1$ and $k=0,1$.
Since $\mathbb{F}_{p^m}$ is a subring of $R=\mathbb{F}_{p^m}+u\mathbb{F}_{p^m}$, $\mathbb{F}_{p^m}[x]$ is a subring of $R[x]$.
Hence we can regard $a(x)$ as a polynomial in $R[x]$. Then $a(x)$ divided by $x^{p^sn}-(\alpha+u\beta)\in R[x]$
gives out a unique remainder
$a_0(x)+\beta^{-1}\left((\alpha+u\beta)-\alpha\right)a_1(x)=a_0(x)+ua_1(x)$.
Therefore,
in $R[x]$ we have
$$a(x)\equiv a_0(x)+ua_1(x)=\sum_{t=0}^{p^sn-1}\gamma_tx^t
\ ({\rm mod} \ x^{p^sn}-(\alpha+u\beta)),$$
where $\gamma_t=a_{0,t}+ua_{1,t}\in R$ for all $t=0,1,\ldots,p^sn-1$.

\par
   As stated above, we see that every element $a(x)$ of $\mathcal{A}$ corresponds to a unique
element $a_0(x)+ua_1(x)=\sum_{t=0}^{p^sn-1}\gamma_tx^t$ of $\mathcal{R}_{\alpha+u\beta}$.
Now,
we define a map $\Phi: \mathcal{A}\rightarrow \mathcal{R}_{\alpha+u\beta}$ by
\begin{equation}
\Phi(a(x))=a(x) \ ({\rm mod} \ x^{p^sn}-(\alpha+u\beta)) \ (\forall a(x)\in \mathcal{A}).
\end{equation}
As $\Phi\left((x^{p^sn}-\alpha)^2\right)=((\alpha+u\beta)-\alpha)^2=\beta^2u^2=0$ in $\mathcal{R}_{\alpha+u\beta}$, we see that
$\Phi$ is well-defined.  Moreover, we have the following

\vskip 3mm \noindent
  {\bf Lemma 2.1} \textit{Using the notations above, the map $\Phi$ is a ring isomorphism from $\mathcal{A}$
onto $\mathcal{R}_{\alpha+u\beta}$}.

\vskip 3mm \noindent
  \textit{Proof} For any $b(x)\in R[x]$, we define $\Phi_0(b(x))=b(x)$ (mod $x^{p^sn}-(\alpha+u\beta)$). Then it is clear
that $\Phi_0$ is a surjective ring homomorphism from $R[x]$ onto $\mathcal{R}_{\alpha+u\beta}=R[x]/\langle x^{p^sn}-(\alpha+u\beta)\rangle$. Since
\begin{eqnarray*}
\Phi_0\left((x^{p^sn}-\alpha)^2\right)&=&(x^{p^sn}-\alpha)^2 \ {\rm mod} \ x^{p^sn}-(\alpha+u\beta)\\
  &=&\left((\alpha+u\beta)-\alpha\right)^2
  =0,
\end{eqnarray*}
by classical ring theory
we see that $\Phi_0$ induces a surjective ring homomorphism from $R[x]/\langle (x^{p^sn}-\alpha)^2\rangle\rangle_{R[x]}$ onto $
\mathcal{R}_{\alpha+u\beta}$.

\par
  Since
$\mathbb{F}_{p^m}[x]$ is a subring of $R[x]$ and $(x^{p^sn}-\alpha)^2\in \mathbb{F}_{p^m}[x]$,
we can regard $\mathcal{A}=\mathbb{F}_{p^m}[x]/\langle (x^{p^sn}-\alpha)^2\rangle_{\mathbb{F}_{p^m}[x]}$ as a subring of $R[x]/\langle (x^{p^sn}-\alpha)^2\rangle_{R[x]}$. Hence the restriction of $\Phi_0$ on $\mathcal{A}$, denoted by $\Phi$ and given by by Equation (1),
is a ring homomorphism from $\mathcal{A}$ to $\mathcal{R}_{\alpha+u\beta}$.

\par
  Moreover, for any $a_0(x)+ua_1(x)\in \mathcal{R}_{\alpha+u\beta}$ where $a_0(x),a_1(x)\in \mathbb{F}_{p^m}[x]$
having degree as most $p^sn-1$, we see that $a_0(x)+\beta^{-1}(x^{p^sn}-\alpha)a_1(x)\in \mathcal{A}$
satisfying $\Phi\left(a_0(x)+\beta^{-1}(x^{p^sn}-\alpha)a_1(x)\right)=a_0(x)+ua_1(x)$. Hence $\Phi$ is surjective.
From this and by $|\mathcal{A}|=p^{2mp^sn}=|\mathcal{R}_{\alpha+u\beta}|$, we deduce that
$\Phi$ is a bijection. Therefore, $\Phi$ is a ring isomorphism from $\mathcal{A}$ onto $\mathcal{R}_{\alpha+u\beta}$.
\hfill $\Box$

 \vskip 3mm \par
  By Lemma 2.1, in order to determine all ideals of $\mathcal{R}_{\alpha+u\beta}$
it is sufficient to determine all ideals of the ring $\mathcal{A}$.

\par
   Since $\alpha\in \mathbb{F}_{p^m}^{\times}$ and
$\mathbb{F}_{p^m}^{\times}$ is a multiplicative cyclic group of order $p^m-1$, there is a unique element
$\alpha_0\in \mathbb{F}_{p^m}^{\times}$ such that $\alpha_0^{p^s}=\alpha$, which implies
$x^{p^sn}-\alpha=(x^n-\alpha_0)^{p^s}$ in $\mathbb{F}_{p^m}[x]$. As ${\rm gcd}(p,n)=1$, there are pairwise coprime monic
irreducible polynomials in $\mathbb{F}_{p^m}[x]$ such that $x^n-\alpha_0=f_1(x)\ldots f_r(x)$, where $f_1(x),\ldots,f_r(x)$ are pairwise coprime monic irreducible polynomials in $\mathbb{F}_{p^m}[x]$, and
${\rm deg}(f_j(x))=d_j$ for all $j=1,\ldots,r$. Then we have
\begin{equation}
(x^{p^sn}-\alpha)^2=(x^n-\alpha_0)^{2p^s}=f_1(x)^{2p^s}\ldots f_r(x)^{2p^s},
\end{equation}
From this one can easily verify the following lemma.

\vskip 3mm
\noindent
  {\bf Lemma 2.2} \textit{$\mathcal{A}=\mathbb{F}_{p^m}[x]/\langle (x^{p^sn}-\alpha)^2\rangle$. Then $\mathcal{A}$ is a principal ideal ring with
$(2p^s+1)^r$-ideals given by}:
$C_{(l_1,\ldots,l_r)}=\langle f_1(x)^{l_1}\ldots f_r(x)^{l_s}\rangle, \ 0\leq l_1,\ldots,l_r\leq 2p^s.$

\vskip 3mm
\par
  By Lemmas 2.1 and 2.2, we see that
all distinct ideals of $\mathcal{R}_{\alpha+u\beta}$ are given by: $\Phi(C_{(l_1,\ldots,l_r)})$, $0\leq l_1,\ldots,l_r\leq 2p^s$.
Next question we concern is that how many codewords contained in $\Phi(C_{(l_1,\ldots,l_r)})$.

\par
   We denote
$F_j(x)=\frac{(x^{p^s}-\alpha)^2}{f_j(x)^{2p^s}}=\prod_{1\leq i\leq r, \ i\neq j}f_i(x)^{2p^s}$ in the rest of this paper. It is clear that ${\rm gcd}(F_j(x),f_j(x)^{2p^s})=1$. So there
exist $g_j(x),h_j(x)\in \mathbb{F}_{p^m}[x]$ such that ${\rm deg}(g_j(x))<{\rm deg}(f_j(x)^{2p^s})=2p^sd_j$ and
\begin{equation}
g_j(x)F_j(x)+h_j(x)f_j(x)^{2p^s}=1.
\end{equation}
From now on, we adopt the following notations.

\vskip 3mm \noindent
  {\bf Notation 2.3} Using the notations above, for any $1\leq j\leq r$ denote
$\mathcal{S}_j=\mathbb{F}_{p^m}[x]/\langle f_j(x)^{2p^s}\rangle$ and let $\theta_j(x)\in \mathcal{A}$ satisfying
\begin{equation}
\theta_j(x)\equiv g_i(x)F_j(x) \ ({\rm mod} \ (x^{p^sn}-\alpha)^2).
\end{equation}

\vskip 3mm \par
  As $f_j(x)$ is a monic irreducible polynomial in $\mathbb{F}_{p^m}[x]$ of degree $d_j$, we have the following lemma.

\vskip 3mm
\noindent
  {\bf Lemma 2.4} (cf. [5] Lemma 3.7 and [6] Example 2.1) \textit{Let $1\leq j\leq r$. Then $\mathcal{S}_j$ satisfies the following properties}:

\vskip 2mm\par
  (i) \textit{$\mathcal{S}_j$ is a finite chain ring, $f_j(x)$ generates the unique
maximal ideal $\langle f_j(x)\rangle$ of $\mathcal{S}_j$, the nilpotency index of $f_j(x)$ is equal to $2p^s$ and the redidue class field of $\mathcal{S}_j$ modulo $\langle f_j(x)\rangle$ is $\mathcal{S}_j/\langle f_j(x)\rangle\cong \mathbb{F}_{p^m}[x]/\langle f_j(x)\rangle$, where $\mathbb{F}_{p^m}[x]/\langle f_j(x)\rangle$ is an extension field of $\mathbb{F}_{p^m}$ with $p^{md_j}$ elements}.

\vskip 2mm\par
  (ii) \textit{Let ${\cal T}_j=\{\sum_{i=0}^{d_j-1}t_ix^i\mid t_0,t_1,\ldots,t_{d_j-1}\in \mathbb{F}_{p^m}\}$. Then we have that ${\cal T}_j=\mathbb{F}_{p^m}[x]/\langle f_j(x)\rangle$ as sets, and every element $\xi$ of $\mathcal{S}_j$ has a unique $f_j(x)$-adic expansion:
$\xi=\sum_{k=0}^{2p^s-1}b_k(x)f(x)^k$, where $b_k(x)\in {\cal T}_j$ for all $k=0,1,\ldots,2p^s-1$.
Moreover, $\xi\in \mathcal{S}_j^{\times}$ if and only if $b_0(x)\neq 0$}.

\vskip 2mm\par
  (iii) \textit{All distinct ideals of $\mathcal{S}_j$ are given by: $\langle f_j(x)^l\rangle=f_j(x)^l\mathcal{S}_j$, $l=0,1,\ldots$, $2p^k$. Moreover, $|\langle f_j(x)^l\rangle|=p^{md_j(2p^s-l)}$ for $l=0,1,\ldots,2p^s$}.

\vskip 3mm \par
  Then by the Chinese remainder theorem for polynomial ring $\mathbb{F}_{p^m}[x]$,
 we have the following conclusions.

\vskip 3mm
\noindent
  {\bf Lemma 2.5} \textit{Using the notations above, we have the following}:

\vskip 2mm\par
  (i) \textit{$\theta_1(x)+\ldots+\theta_r(x)=1$, $\theta_j(x)^2=\theta_j(x)$
and $\theta_j(x)\theta_l(x)=0$  in the ring $\mathcal{A}$ for all $1\leq j\neq l\leq r$}.

\vskip 2mm\par
  (ii) \textit{For any $a_j(x)\in \mathcal{S}_j$ for $j=1,\ldots,r$, define
\begin{center}
$\Psi(a_1(x),\ldots,a_r(x))=\sum_{j=1}^r\theta_j(x)a_j(x)$ $(${\rm mod} $(x^{p^kn}-\alpha)^2)$.
\end{center}
Then
$\Psi$ is a ring isomorphism from $\mathcal{S}_1\times\ldots\times\mathcal{S}_r$ onto $\mathcal{A}$, and the inverse
$\Psi^{-1}: \mathcal{A}\rightarrow \mathcal{S}_1\times\ldots\times\mathcal{S}_r$ of $\Psi$ is given by}:
$$\Psi^{-1}(b(x))=\left(b(x) \ ({\rm mod} \ f_1(x)^{2p^k}), \ldots, b(x) \ ({\rm mod} \ f_r(x)^{2p^k})\right) \ (\forall b(x)\in \mathcal{A}).$$

\vskip 3mm \par
  Now, we can list all distinct $(\alpha+u\beta)$-constacyclic codes over $R$ of length $p^sn$ by the following theorem.

\vskip 3mm \noindent
   {\bf Theorem 2.6} \textit{Using the notations above, all distinct $(\alpha+u\beta)$-constacyclic
codes over $R$ of length $p^sn$ are given by}:
$$\mathcal{C}_{(l_1,\ldots,l_r)}=\langle f_1(x)^{l_1}\ldots f_r(x)^{l_r}\rangle \ ({\rm mod} \ x^{p^sn}-(\alpha+u\beta)),$$
\textit{where $0\leq l_1,\ldots,l_r\leq 2p^s$, and the number of codewords in $\mathcal{C}_{(l_1,\ldots,l_r)}$ is equal to}
$|\mathcal{C}_{(l_1,\ldots,l_r)}|=p^{m(2p^sn-\sum_{j=1}^rd_jl_j)}.$
 \textit{Therefore, the number of $(\alpha+u\beta)$-constacyclic
codes over $R$ of length $p^sn$ is equal to $(2p^s+1)^r$}.

\vskip 3mm \noindent
   \textit{Proof} By lemmas 2.1 and 2.2, we see that $(\alpha+u\beta)$-constacyclic codes over $R$ of length $p^sn$, i.e., all distinct ideals
of $\mathcal{R}_{\alpha+u\beta}$, are given by:
$$
\mathcal{C}_{(l_1,\ldots,l_r)}=\Phi\left(C_{(l_1,\ldots,l_r)}\right)
=C_{(l_1,\ldots,l_r)} \ ({\rm mod} \ x^{p^sn}-(\alpha+u\beta)),
$$
where $C_{(l_1,\ldots,l_r)}=\langle f_1(x)^{l_1}\ldots f_r(x)^{l_s}\rangle$ is the ideal of $\mathcal{A}$ generated by $f_1(x)^{l_1}$ $\ldots f_r(x)^{l_s}$, $0\leq l_1,\ldots,l_r\leq 2p^s$. Hence the number of $(\alpha+u\beta)$-constacyclic
codes over $R$ of length $p^sn$ is equal to $(2p^s+1)^r$ by Lemma 2.2.

\par
  For any integer $j$, $1\leq j\leq r$, by $F_j(x)=\prod_{1\leq k\leq r, \ k\neq j}f_k(x)^{2p^s}$
and Equation (4) it follows that
\begin{equation}
 g_j(x)w_j(x)\left(\prod_{1\leq k\leq r, \ k\neq j}f_j(x)^{l_j}\right)+h_j(x)f_j(x)^{2p^s}=1
\end{equation}
where $w_j(x)=\frac{F_j(x)}{\prod_{1\leq k\leq r, \ k\neq j}f_k(x)^{l_k}}\in\mathbb{F}_{p^m}[x]$.
Now, we select $\delta_j\in \mathcal{S}_j$ satisfying
$\delta_j\equiv \prod_{1\leq k\leq r, \ k\neq j}f_k(x)^{l_k}$ (mod $f_j(x)^{2p^s}$) as polynomials. By (5) it follows that
$(g_j(x)w_j(x))\delta_j\equiv 1$  (mod $f_j(x)^{2p^s}$).
Hence $\delta_j$ is an invertible element of $\mathcal{S}_j$. Therefore,
\begin{eqnarray*}
&&C_{(l_1,\ldots,l_r)} \ ({\rm mod} \ f_j(x)^{2p^s})
  =\langle f_1(x)^{l_1}\ldots f_r(x)^{l_r} ({\rm mod} \ f_j(x)^{2p^s})\ \rangle_{\mathcal{S}_j}\\
  &=& \langle f_j(x)^{l_j}\prod_{1\leq k\leq r, \ k\neq j}f_k(x)^{l_k} \ ({\rm mod} \ f_j(x)^{2p^s})\ \rangle_{\mathcal{S}_j}\\
  &=& \langle \delta_jf_j(x)^{l_j}\rangle_{\mathcal{S}_j}
  = \langle f_j(x)^{l_j}\rangle.
\end{eqnarray*}
From this and by Lemma 2.5(ii), we deduce that $\Psi^{-1}(C_{(l_1,\ldots,l_r)})
=\langle f_1(x)^{l_1}\rangle\times\ldots\times\langle f_r(x)^{l_r}\rangle$. Then by Lemmas 2.1 and 2.4, it follows that
\begin{eqnarray*}
|\mathcal{C}_{(l_1,\ldots,l_r)}|&=&|C_{(l_1,\ldots,l_r)}|=\prod_{j=1}^r|\langle f_j(x)^{l_j}\rangle|=\prod_{j=1}^rp^{md_j(2p^s-l_j)}\\
  &=&p^{m(2ps\sum_{j=1}^rd_j-\sum_{j=1}^rd_jl_j)}=p^{m(2p^sn-\sum_{j=1}^rd_jl_j)}.
\end{eqnarray*}
\hfill $\Box$

\vskip 3mm \par
  Using the notations of Theorem 2.6, let $g_{(l_1,\ldots,l_r)}(x)\in \mathcal{R}_{(\alpha+u\beta)}$
satisfying $$g_{(l_1,\ldots,l_r)}(x)\equiv f_1(x)^{l_1}\ldots f_r(x)^{l_r} \ ({\rm mod} \ x^{p^sn}-(\alpha+u\beta)).$$
Then $g_{(l_1,\ldots,l_r)}(x)$ is called the \textit{generator} of the $(\alpha+u\beta)$-constacyclic code $\mathcal{C}_{(l_1,\ldots,l_r)}$.

\par
   For any $\xi=(\xi_0,\xi_1,\ldots,\xi_{p^sn-1})\in R^{p^sn}$, where $\xi_i=a_i+b_iu\in R=\mathbb{F}_{p^m}+u\mathbb{F}_{p^m}$ with
$a_i,b_i\in \mathbb{F}_{p^m}$ for $i=0,1,\ldots,p^sn-1$,
recall that the Hamming weight ${\rm wt}_H(\xi)$ of $\xi$ is defined by
${\rm wt}_H(\xi)=|\{i\mid
\xi_i\neq 0, \ 0\leq i\leq p^sn-1\}|=|\{i\mid a_i\neq 0 \ {\rm or} \ b_i\neq 0, \ 0\leq i\leq p^sn-1\}|$.
For any linear code $\mathcal{C}$ over $R$, the minimum
distance $d(\mathcal{C}$ of $\mathcal{C}$ is equal to
$d(\mathcal{C})={\rm min}\{{\rm wt}_H(\xi)\mid \xi\neq (0,0,\ldots,0), \ \xi\in \mathcal{C}\}.$

\par
    In order to determine the minimum
distance $d_{(l_1,\ldots,l_r)}$ of $\mathcal{C}_{(l_1,\ldots,l_r)}$, we need the following lemma which is inferred from
[14] Theorem 4.2.

\vskip 3mm \noindent
   {\bf Lemma 2.7} \textit{Let $\mathcal{C}$ be a linear code over $R$ of length $N$. Then $d(\mathcal{C})=d(\overline{(\mathcal{C}:u)})$,
where $\overline{(\mathcal{C}:u)}=\{a\mid u(a+ub)\in \mathcal{C}, \ a,b\in \mathbb{F}_{p^m}^N\}=\{a\in \mathbb{F}_{p^m}\mid ua\in \mathcal{C}\}$}.

\vskip 3mm \noindent
   {\bf Theorem 2.8} \textit{Let
$\mathcal{C}_{(l_1,\ldots,l_r)}$ be an $(\alpha+u\beta)$-constacyclic code over $R$ of length $p^sn$ given
in Theorem 2.6 and has generator $g_{(l_1,\ldots,l_r)}(x)=\sum_{i=0}^{p^sn-1}g_ix^i$, where $g_i\in R$ for all $i=0,1,\ldots,p^sn-1$. Decompose}
$$\left(\begin{array}{ccccc}g_0& g_1& \ldots & g_{p^sn-2}& g_{p^sn-1}\cr
(\alpha+u\beta)g_{p^sn-1} &g_1& \ldots & g_{p^sn-3}& g_{p^sn-2}
\cr \ldots &\ldots &\ldots &\ldots &\ldots \cr
(\alpha+u\beta)g_{2} &(\alpha+u\beta)g_{3}& \ldots & g_{0}& g_{1}\cr
(\alpha+u\beta)g_{1} &(\alpha+u\beta)g_{1}& \ldots & (\alpha+u\beta)g_{1}& g_{0}
\end{array}\right)$$

\noindent
\textit{into $G_0+uG_1$ where $G_0$ and $G_1$ are $p^sn\times p^sn$ matrices over $\mathbb{F}_{p^m}$, and
set}

\noindent
   $\bullet$  \textit{$C_t=\{aG_t\mid a=(a_0,a_1,\ldots,a_{p^sn-1})\in \mathbb{F}_{p^m}^{p^sn}\}$, $t=0,1$};

\noindent
   $\bullet$ \textit{ $C=\{bG_0+aG_1\mid aG_0=0, \ a,b\in \mathbb{F}_{p^m}^{p^sn}\}$}.

\noindent
 \textit{Then we have the following one of the following cases}:

\vskip 2mm \par
   (i) \textit{When $G_0\neq 0$, $d(\mathcal{C}_{(l_1,\ldots,l_r)})=d(C)\leq d(C_0)$. Especially,
$d(\mathcal{C}_{(l_1,\ldots,l_r)})=d(C_0)$ if $aG_1=0$ for any $a\in \mathbb{F}_{p^m}^{p^sn}$ satisfying $aG_0=0$}.

\vskip 2mm \par
   (ii) \textit{When $G_0=0$, $d(\mathcal{C}_{(l_1,\ldots,l_r)})=d(C_1)$}.

\vskip 3mm \noindent
   \textit{Proof} Denote $\mathcal{C}=\mathcal{C}_{(l_1,\ldots,l_r)}$. It is clear that $G_0+uG_1$ is a generator matrix of $\mathcal{C}$ as a
linear code over $R$. Hence $\mathcal{C}=\{(a+ub)(G_0+uG_1)\mid a,b\in \mathbb{F}_{p^m}^{p^sn}\}
           =\{aG_0+u(bG_0+aG_1)\mid a,b\in \mathbb{F}_{p^m}^{p^sn}\}$.
Let $w\in \mathbb{F}_{p^m}^{p^sn}$. Then $uw\in \mathcal{C}$ if and only if
there exist $a,b\in \mathbb{F}_{p^m}^{p^sn}$ such that $uw=aG_0+u(bG_0+aG_1)$, which is equivalent
that $aG_0=0$ and $w=bG_0+aG_1$. Therefore,
$$\overline{(\mathcal{C}:u)}=\{bG_0+aG_1\mid aG_0=0, \ a,b\in \mathbb{F}_{p^m}^{p^sn}\}=C.$$
By Lemma 2.7, we have $d(\mathcal{C})=d(C)$.

\par
  (i) By $C_0\subseteq C$ (set $a=0$ in $C$), we have $d(C)\leq d(C_0)$.
If $aG_1=0$ for any $a\in \mathbb{F}_{p^m}^{p^sn}$ satisfying $aG_0=0$, we have $C=C_0$, and
hence $d(C)=d(C_0)$.

\par
  (ii) When $G_0=0$, we have $C=C_1$, and
hence $d(C)=d(C_1)$.
\hfill $\Box$

\vskip 3mm \par
  Finally, we consider the special situation of $r=1$.

\vskip 3mm \noindent
   {\bf Lemma 2.9} (cf. [16] Theorem 10.7) \textit{Let $n\geq 2$ be an integer.
Let $\alpha_0\in \mathbb{F}_{p^m}^{\times}$ and ${\rm ord}(\alpha_0)=\kappa>1$ in $\mathbb{F}_{p^m}^{\times}$. Then
the binomial $x^n-\alpha_0\in \mathbb{F}_{p^m}[x]$ is irreducible over $\mathbb{F}_{p^m}$ if and only if the following two conditions are satisfied}:

\vskip 2mm \par
  (I) \textit{Every prime divisor of $n$ divides $\kappa$, but not $(p^m-1)/\kappa$}.

\vskip 2mm \par
  (II) \textit{If $4|n$ then $4|(p^m-1)$}.

\vskip 3mm \par
   Then by Theorem 2.6, we have the following corollary.

\vskip 3mm \noindent
   {\bf Corollary 2.10}  \textit{Let $\alpha=\alpha_0^{p^s}$ and $\alpha_0,\beta\in \mathbb{F}_{p^m}^{\times}$ such that $\alpha_0$ satisfies
Conditions (I) and (II) in lemma 2.7.
Then all $(2p^s+1)$ distinct $(\alpha+u\beta)$-constacyclic codes over $R$ of length $p^sn$ are given by}:

\vskip 2mm\noindent
 \textit{$\mathcal{C}_l=\langle (x^n-\alpha_0)^l\rangle$ if $0\leq l\leq p^s-1$};
\textit{$\mathcal{C}_l=\langle u(x^n-\alpha_0)^{l-p^s}\rangle$ if $p^s\leq l\leq 2p^s$}.

\vskip 2mm\noindent
  \textit{Moreover, the number of codewords in $\mathcal{C}_l$ is equal to $|\mathcal{C}_l|=p^{mn(2p^s-l)}$}.

\vskip 3mm \noindent
   \textit{Proof} By Lemma 2.7, $x^n-\alpha_0$ is an irreducible polynomial in $\mathbb{F}_{p^m}[x]$.
Hence $(x^{p^sn}-\alpha)=(x^n-\alpha_0)^{2p^s}$. From this and by Theorem 2.6, we deduce that all distinct
$(\alpha+u\beta)$-constacyclic codes over $\mathbb{F}_{p^m}+u\mathbb{F}_{p^m}$ of length $p^sn$ are given by:
$\mathcal{C}_l=\langle (x^n-\alpha_0)^l\rangle$ (mod $x^{p^sn}-(\alpha+u\beta)$) with
$|\mathcal{C}_l|=p^{mn(2p^s-l)}$, $0\leq l\leq 2p^s$. Then the conclusions
follow from $(x^n-\alpha_0)^{p^s}=x^{p^sn}-\alpha\equiv (\alpha+u\beta)-\alpha=u\beta$  (mod $x^{p^sn}-(\alpha+u\beta)$)
and $\beta\in \mathbb{F}_{p^m}^{\times}$, immediately.
\hfill $\Box$


\section{Dual codes of $(\alpha+u\beta)$-constacyclic codes over $R$} \label{}
\noindent
In this section,
we consider the dual code of any $(\alpha+u\beta)$-constacyclic code over $R$ of length $p^sn$.

\par
  Let $\mathcal{C}_{(l_1,\ldots,l_r)}$ be an arbitrary $(\alpha+u\beta)$-constacyclic code over $R$ of length $p^sn$ given by Theorem 2.6. Then by Lemma 1.1, we know that
$\mathcal{C}_{(l_1,\ldots,l_r)}^{\bot_E}$ must be an ideal of the ring $R[x]/\langle x^{p^sn}-(\alpha+u\beta)^{-1}\rangle$, where
$(\alpha+u\beta)^{-1}=\alpha^{-1}+u(-\alpha^{-2}\beta)$. In order to simplify the notations, in the following we denote

\vskip 2mm\noindent
  $\bullet$ $\mathcal{R}_{(\alpha+u\beta)^{-1}}=R[x]/\langle x^{p^sn}-(\alpha+u\beta)^{-1}\rangle$ and $\widehat{\mathcal{A}}=\mathbb{F}_{p^m}[x]/\langle (x^{p^sn}-\alpha^{-1})^2\rangle$.

\vskip 2mm\par
  For any polynomial $f(x)=\sum_{i=0}^dc_ix^i\in \mathbb{F}_{p^m}[x]$ of degree $d\geq 1$, recall that
the \textit{reciprocal polynomial} of $f(x)$ is defined as $\widetilde{f}(x)=\widetilde{f(x)}=x^df(\frac{1}{x})=\sum_{i=0}^dc_ix^{d-i}$, and
 $f(x)$ is said to be \textit{self-reciprocal} if $\widetilde{f}(x)=\delta f(x)$ for some $\delta\in \mathbb{F}_{p^m}^{\times}$.
It is known that $\widetilde{\widetilde{f}(x)}=f(x)$ if $f(0)\neq 0$, and $\widetilde{f(x)g(x)}=\widetilde{f}(x)\widetilde{g}(x)$ for
any monic polynomials $f(x), g(x)\in\mathbb{F}_{p^m}[x]$ with positive degrees satisfying $f(0),g(0)\in \mathbb{F}_{p^m}^{\times}$.
 Then by Equation (2) in Section 3 and $x^{p^sn}-\alpha^{-1}=-\alpha^{-1}(1-\alpha x^{p^sn})=-\alpha^{-1}\widetilde{(x^{p^sn}-\alpha)}$, it follows
that
$$x^{p^sn}-\alpha^{-1}=-\alpha^{-1}\widetilde{f}_1(x)^{p^s}\ldots \widetilde{f}_r(x)^{p^s}.$$
Since $f_1(x),\ldots,f_r(x)$ are pairwise coprime monic basic irreducible polynomials in $\mathbb{F}_{p^m}[x]$,
$\widetilde{f}_1(x),\ldots, \widetilde{f}_r(x)$  are pairwise coprime basic irreducible polynomials in $\mathbb{F}_{p^m}[x]$ as well. Moreover, ${\rm deg}(\widetilde{f}_j(x))={\rm deg}(f_j(x))=d_j$ for all $j=1,\ldots,r$. From these, similar to Lemma 2.2 we deduce that
all distinct ideals of the ring $\widehat{\mathcal{A}}$ are given by:
$\langle \widetilde{f}_1(x)^{t_1}\ldots \widetilde{f}_r(x)^{t_r}\rangle,
\ 0\leq t_1,\ldots,t_r\leq 2p^s.$
Moreover, we have
$|\langle \widetilde{f}_1(x)^{t_1}\ldots \widetilde{f}_r(x)^{t_r}\rangle|=p^{m(2p^sn-\sum_{j=1}^rd_jt_j)}$.

\par
  For any $a(x)\in \widehat{\mathcal{A}}$,  $a(x)$ can be uniquely expressed as
a polynomial in $\mathbb{F}_{p^m}[x]$ has degree at most $2p^sn-1$. We define
$$\widehat{\Phi}(a(x))=a(x) \ ({\rm mod} \ x^{p^sn}-(\alpha+u\beta)^{-1}).$$
Then similar to Lemma 2.1 and its proof, we conclude $\widehat{\Phi}$ is a ring isomorphism
from $\widehat{\mathcal{A}}$ onto $\mathcal{R}_{(\alpha+u\beta)^{-1}}$. Therefore, all distinct $(\alpha+u\beta)^{-1}$-constacyclic
codes over $R$ of length $p^sn$, i.e., all distinct ideals of $\mathcal{R}_{(\alpha+u\beta)^{-1}}$ are given by:
\begin{eqnarray*}
\widehat{\mathcal{C}}_{(t_1,\ldots,t_r)}&=&\widehat{\Phi}\left(\langle \widetilde{f}_1(x)^{t_1}\ldots \widetilde{f}_r(x)^{t_r}\rangle\right)\\
  &=& \langle \widetilde{f}_1(x)^{t_1}\ldots \widetilde{f}_r(x)^{t_r}\rangle \ ({\rm mod} \ x^{p^sn}-(\alpha+u\beta)^{-1}),
\end{eqnarray*}
where $0\leq t_1,\ldots,t_r\leq 2p^s$. Moreover, we have $|\widehat{\mathcal{C}}_{(t_1,\ldots,t_r)}|=p^{m(2p^sn-\sum_{j=1}^rd_jt_j)}$.

\par
  In order to obtain the dual code of any $(\alpha+u\beta)$-constacyclic code over $R$ of length $p^sn$ we
need to build a relationship between $\mathcal{R}_{\alpha+u\beta}$ and $\mathcal{R}_{(\alpha+u\beta)^{-1}}$. To do this, we
define
\begin{equation}
\tau: \mathcal{A}\rightarrow \widehat{\mathcal{A}}
\ {\rm via} \ \tau: c(x)\mapsto c(x^{-1}) \ (\forall c(x)\in \mathcal{A}).
\end{equation}

\vskip 3mm \noindent
   {\bf Lemma 3.1} \textit{The map $\tau$ defined by $(6)$ is a ring isomorphism from
$\mathcal{A}$ onto $\widehat{\mathcal{A}}$}.

\vskip 3mm \noindent
   \textit{Proof} We define a map $\tau_0: \mathbb{F}_{p^m}[x]\rightarrow \widehat{\mathcal{A}}$ by
$\tau_0(\rho(x))=\rho(x^{-1})=\sum_{i}\rho_ix^{-i}$
for any $\rho(x)=\sum_{i}\rho_ix^i\in \mathbb{F}_{p^m}[x]$
with  $\rho_i\in \mathbb{F}_{p^m}$, where $x^{-1}\in \widehat{\mathcal{A}}$. It is clear that $\tau_0$ is a ring homomorphism
from $\mathbb{F}_{p^m}[x]$ to $\widehat{\mathcal{A}}$. Since $(x^{p^sn}-\alpha^{-1})^2=0$ in $\widehat{\mathcal{A}}$, we have
$x^{2p^sn}-2\alpha^{-1}x^{p^sn}+\alpha^{-2}=0$, which implies
\begin{equation}
x^{-1}=\alpha^2x^{p^sn-1}(2\alpha^{-1}-x^{p^sn})\in \widehat{\mathcal{A}}^{\times}.
\end{equation}

\par
   For any $\eta(x)=\sum_{k=0}^{2p^sn-1}\eta_kx^k\in \widehat{\mathcal{A}}$ with $\eta_k\in \mathbb{F}_{p^m}$, we select
$\rho(x)=\sum_{k=0}^{2p^sn-1}\eta_k\left(\alpha^2x^{p^sn-1}(2\alpha^{-1}-x^{p^sn})\right)^k \in \mathbb{F}_{p^m}[x]$. Then by
(7) it follows that
$$\tau_0(\rho(x))=\sum_{k=0}^{2p^sn-1}\eta_k\left(\alpha^2x^{-(p^sn-1)}(2\alpha^{-1}-x^{-p^sn})\right)^k
=\eta(x).$$
Therefore, $\tau_0$ is a surjective ring homomorphism
from $\mathbb{F}_{p^m}[x]$ onto $\widehat{\mathcal{A}}$.

\par
  Moreover, by $\tau_0((x^{p^sn}-\alpha)^2)=(x^{-p^sn}-\alpha)^2=\alpha^2x^{-2p^sn}(x^{p^sn}-\alpha^{-1})^2=0$ in $\widehat{\mathcal{A}}$
and classical ring theorem, we deduce that $\tau_0$ induces a surjective ring homomorphism $\tau$ from
$\mathcal{A}=\mathbb{F}_{p^m}[x]/\langle (x^{p^sn}-\alpha)^2\rangle$ onto $\widehat{\mathcal{A}}$ defined by (6).
But $|\mathcal{A}|=p^{2mp^sn}=|\widehat{\mathcal{A}}|$, we conclude that
$\tau$ is a bijection, and hence $\tau$ a ring isomorphism from $\mathcal{A}$ onto $\widehat{\mathcal{A}}$.
\hfill $\Box$

\vskip 3mm\par
   Now, we denote $\overline{\tau}=\widehat{\Phi}\tau\Phi^{-1}$. Then by Lemma 2,1, Lemma 3.1 and the properties
of $\widehat{\Phi}$, we see that $\overline{\tau}$ is a ring isomorphism from $\mathcal{R}_{\alpha+u\beta}$
onto $\mathcal{R}_{(\alpha+u\beta)^{-1}}$ such that the diagram
{\small $\begin{array}{ccc} \mathcal{A} & \stackrel{\tau}{\longrightarrow} &  \widehat{\mathcal{A}} \cr
  \Phi \downarrow &  & \ \ \ \downarrow \widehat{\Phi} \cr
\mathcal{R}_{\alpha+u\beta} & \stackrel{\overline{\tau}}{\longrightarrow} &  \mathcal{R}_{(\alpha+u\beta)^{-1}}.
\end{array}$} commutes.
In the ring $\mathcal{R}_{\alpha+u\beta}$, we have $x^{p^sn}-(\alpha+u\beta)=0$, which implies
$u=\varrho(x)$ where $\varrho(x)=\beta^{-1}(x^{p^sn}-\alpha)$. Now, we regard $\varrho(x)$ as an element of $\mathcal{A}$. Then by Equation (1) in Section 2 it follows that
$\Phi(\varrho(x))=u\equiv \beta^{-1}(x^{p^sn}-\alpha) \ ({\rm mod} \ x^{p^sn}-(\alpha+u\beta)),$
which implies $\Phi^{-1}(u)=\varrho(x)\in \mathcal{A}$. From this and by (6), we deduce that
\begin{eqnarray*}
\overline{\tau}(u)&=&\widehat{\Phi}\tau(\Phi^{-1}(u))=\widehat{\Phi}(\tau(\varrho(x)))
  =\widehat{\Phi}(\varrho(x^{-1}))=\widehat{\Phi}(\beta^{-1}(x^{-p^sn}-\alpha))\\
  &=&\beta^{-1}(x^{-p^sn}-\alpha) \ ({\rm mod} \ x^{p^sn}-(\alpha+u\beta)^{-1})\\
  &=&\beta^{-1}((\alpha+u\beta)-\alpha)=u.
\end{eqnarray*}
Therefore, the ring isomorphism $\overline{\tau}: \mathcal{R}_{\alpha+u\beta}\rightarrow \mathcal{R}_{(\alpha+u\beta)^{-1}}$ is given by:
\begin{center}
$\overline{\tau}(c(x))=c(x^{-1})=\sum_{i=0}^{p^sn-1}c_ix^{-i}$,
\end{center}
for any $c(x)=\sum_{i=0}^{p^sn-1}c_ix^i\in \mathcal{R}_{\alpha+u\beta}$
with $c_0,c_1,\ldots$, $c_{p^sn-1}\in R$.
Now,  let $a=(a_0,a_1,\ldots,a_{p^sn-1}), b=(b_0,b_1,\ldots,b_{p^sn-1})\in R^{p^sn}$. We define
$$a(x)=\sum_{i=0}^{p^sn-1}a_ix^i\in \mathcal{R}_{\alpha+u\beta} \ {\rm and} \
b(x)=\sum_{i=0}^{p^sn-1}b_ix^i\in \mathcal{R}_{(\alpha+u\beta)^{-1}}.$$
Using these notations, we give the following lemma.

\vskip 3mm \noindent
   {\bf Lemma 3.2} \textit{For any $a,b\in R^{p^sn}$, we have $[a,b]_E=\sum_{i=0}^{p^sn-1}a_ib_i=0$ if
$\overline{\tau}(a(x))\cdot b(x)=0$ in $\mathcal{R}_{(\alpha+u\beta)^{-1}}$}.

\vskip 3mm \noindent
   \textit{Proof} By $x^{p^sn}=(\alpha+u\beta)^{-1}$ and (8), we have
$$\overline{\tau}(a(x))=\sum_{i=0}^{p^sn-1}a_ix^{-i}=x^{-p^sn}\sum_{i=0}^{p^sn-1}a_ix^{p^sn-i}
 =(\alpha+u\beta)\sum_{i=0}^{p^sn-1}a_ix^{p^sn-i}.$$
Then $\overline{\tau}(a(x))\cdot b(x)=[a,b]_E+c_1x+\ldots+c_{p^sn-1}x^{p^sn-1}$ in $\mathcal{R}_{(\alpha+u\beta)^{-1}}$ for
some $c_1,\ldots,c_{p^sn-1}\in R$.
Hence $[a,b]_E=0$, when $\overline{\tau}(a(x))\cdot b(x)=0$.
\hfill $\Box$

\vskip 3mm \par
  Now, we can give the dual code for every $(\alpha+u\beta)$-constacyclic code over $R$ by the following theorem.

\vskip 3mm \noindent
   {\bf Theorem 3.3} \textit{Let $\mathcal{C}_{(l_1,\ldots,l_r)}$
be an $(\alpha+u\beta)$-constacyclic code over $R$ of length $p^sn$ listed in Theorem 2.6.
The dual code $\mathcal{C}_{(l_1,\ldots,l_r)}^{\bot_E}$ of $\mathcal{C}_{(l_1,\ldots,l_r)}$ is given by}
$$\mathcal{C}_{(l_1,\ldots,l_r)}^{\bot_E}=\langle \widetilde{f}_1(x)^{2p^s-l_1}\ldots \widetilde{f}_r(x)^{2p^s-l_r}\rangle \
({\rm mod} \ x^{p^sn}-(\alpha+u\beta)^{-1})$$
\textit{which is an $(\alpha+u\beta)^{-1}$-constacyclic code over $R$ of length $p^sn$}.

\vskip 3mm \noindent
  \textit{Proof} For any ideal $\mathcal{C}=\mathcal{C}_{(l_1,\ldots,l_r)}=\Phi(\langle f_1(x)^{l_1}\ldots f_r(x)^{l_r}\rangle$ of $\mathcal{R}_{\alpha+u\beta}$,
let $\mathcal{D}=\widehat{\Phi}\left(\langle \widetilde{f}_1(x)^{2p^s-l_1}\ldots \widetilde{f}_r(x)^{2p^s-l_r}\rangle\right)$ which
is an ideal of $\mathcal{R}_{(\alpha+u\beta)^{-1}}$. Then
$\mathcal{D}=\langle \widetilde{f}_1(x)^{2p^s-l_1}\ldots \widetilde{f}_r(x)^{2p^s-l_r}\rangle$ (mod $ x^{p^sn}-(\alpha+u\beta)^{-1}$) by
the definition of $\widehat{\Phi}$.
  By $\overline{\tau}=\widehat{\Phi}\tau\Phi^{-1}$, i.e., $\overline{\tau}\Phi=\widehat{\Phi}\tau$, and Theorem 2.6, we have
\begin{eqnarray*}
\overline{\tau}(\mathcal{C})\cdot \mathcal{D}&=&\overline{\tau}\Phi\left(\langle f_1(x)^{l_1}\ldots f_r(x)^{l_r}\rangle\right)\cdot \mathcal{D}
   =\widehat{\Phi}\tau\left(\langle f_1(x)^{l_1}\ldots f_r(x)^{l_r}\rangle\right)\cdot \mathcal{D}\\
  &=&\widehat{\Phi}\left(\langle f_1(x^{-1})^{l_1}\ldots f_r(x^{-1})^{l_r}\rangle\right)\cdot \mathcal{D}\\
  &=&\widehat{\Phi}\left(\langle x^{-\sum_{j=1}^rd_jl_j}\prod_{j=1}^rx^{d_jl_j}f_j(x^{-1})^{l_j}\rangle\right)\cdot \mathcal{D}\\
    &=&\widehat{\Phi}\left(\langle (x^{d_1l_1}f_1(x^{-1})^{l_1})\ldots (x^{d_rl_r}f_r(x^{-1})^{l_r})\rangle\right)\cdot \mathcal{D}\\
  &=&\widehat{\Phi}\left(\langle \prod_{j=1}^r\widetilde{f}_j(x)^{l_j}\rangle\right)
   \cdot \widehat{\Phi}\left(\langle \prod_{j=1}^r\widetilde{f}_j(x)^{2p^s-l_j}\rangle\right)\\
     &=&\widehat{\Phi}\left(\langle \prod_{j=1}^r\widetilde{f}_j(x)^{l_j}\rangle\cdot\langle \prod_{j=1}^r\widetilde{f}_j(x)^{2p^s-l_j}\rangle\right)
\end{eqnarray*}
\begin{eqnarray*}
  &=&\widehat{\Phi}\left(\langle \widetilde{f}_1(x)^{2p^s}\ldots \widetilde{f}_r(x)^{2p^s}\rangle\right)\\
  &=&\widehat{\Phi}\left(\langle(x^{p^sn}-\alpha^{-1})^2\rangle\right)\\
  &=&\{0\},
\end{eqnarray*}
which implies that $\mathcal{D}\subseteq \mathcal{D}^{\bot_E}$ by Lemma 3.2. From this and by
$|\mathcal{C}||\mathcal{D}|=p^{m(2p^sn-\sum_{j=1}^rd_jl_j)}p^{m(2p^sn-\sum_{j=1}^rd_j(2p^s-l_j))}=p^{2mp^sn}=|R|^{p^sn}.$
Since $R$ is a Frobenius ring, we conclude that $\mathcal{D}^{\bot_E}=\mathcal{D}$ (cf. [11]).
\hfill $\Box$

\vskip 3mm\noindent
 {\bf Example 3.4} Let $t$ be a positive integer. We consider $(2+u)$-constacyclic codes of length $6\cdot 5^t$
over $\mathbb{F}_{3}+u\mathbb{F}_3$ and their dual codes. In this case, we have $p=3$, $m=s=1$,
$n=2\cdot 5^t$, $\alpha=2=-1$ and $\beta=1$.
  By [3] Section 4, we know that
$x^{2\cdot 5^t}-2=x^{2\cdot 5^t}+1=f_1(x)\prod_{i=2}^{t+1}f_i(x)\widetilde{f}_i(x)$
is the factorization $x^{2\cdot 5^t}+1$ into monic irreducible factors in $\mathbb{F}_3[x]$, where

\vskip 2mm\par
  $f_1(x)=x^2+1=\widetilde{f}_1(x)$; $f_i(x)=x^{4\cdot 5^{i-2}}+x^{3\cdot 5^{i-2}}+2x^{5^{i-2}}+1$,

\vskip 2mm\par
   $\widetilde{f}_i(x)=x^{4\cdot 5^{i-2}}+2x^{3\cdot 5^{i-2}}+x^{5^{i-2}}+1$, $i=2,\ldots,t+1$.

\vskip 2mm\noindent
  By Theorem 2.6, all $(2+u)$-constacyclic codes of length $6\cdot 5^t$ over $\mathbb{F}_{3}+u\mathbb{F}_3$ are given by:
$$\mathcal{C}_{(l_1,l_2,l^{\prime}_2,\ldots,l_{t+1},l^{\prime}_{t+1})}
=\langle f_1(x)^{l_1}f_2(x)^{l_2}\widetilde{f}_2(x)^{l^{\prime}_2}\ldots
f_t(x)^{l_{t+1}}\widetilde{f}_t(x)^{l^{\prime}_{t+1}}\rangle$$
(mod $ x^{6\cdot 5^t}-(2+u)$), $0\leq l_1,l_2,l^{\prime}_2,\ldots,l_{t+1},l^{\prime}_{t+1}\leq 6$. Moreover, the number
of codewords contained in $\mathcal{C}_{(l_1,l_2,l^{\prime}_2,\ldots,l_{t+1},l^{\prime}_{t+1})}$ is equal to
$|\mathcal{C}_{(l_1,l_2,l^{\prime}_2,\ldots,l_{t+1},l^{\prime}_{t+1})}|=3^{12\cdot 5^t-2l_1-4\sum_{i=2}^{t+1}(l_i+l^{\prime}_i)5^{i-2}}$
and the number of $(2+u)$-constacyclic codes over $\mathbb{F}_{3}+u\mathbb{F}_3$ of length $6\cdot 5^t$ is equal to $7^{2t+1}$.

\par
   As $(2+u)^{-1}=2+2u$ in $\mathbb{F}_{3}+u\mathbb{F}_3$, by Theorem 3.3 we know that
the dual code $\mathcal{C}_{(l_1,l_2,l^{\prime}_2,\ldots,l_{t+1},l^{\prime}_{t+1})}^{\bot_E}$
of $\mathcal{C}_{(l_1,l_2,l^{\prime}_2,\ldots,l_{t+1},l^{\prime}_{t+1})}$ is a $(2+2u)$-constacyclic
cyclic code over $\mathbb{F}_{3}+u\mathbb{F}_3$ of length
$6\cdot 5^t$. Precisely, we have
$$\mathcal{C}_{(l_1,l_2,l^{\prime}_2,\ldots,l_{t+1},l^{\prime}_{t+1})}^{\bot_E}
=\langle f_1(x)^{6-l_1}\prod_{i=2}^{t+1}f_i(x)^{6-l^{\prime}_i}\widetilde{f}_i(x)^{6-l_i}\rangle \
({\rm mod} \ x^{6\cdot 5^t}-(2+2u)).$$

\vskip 3mm \noindent
   {\bf Remark}  Using conclusions in [3], one can determine all distinct $(u\beta-1)$-constacyclic codes
over $R=\mathbb{F}_{p^m}+u\mathbb{F}_{p^m}$ of length $2p^sl^t$ and their dual codes, for any odd prime $l$ coprime to $p$ and positive integer
$t$.

\par
   Let $\lambda=\alpha+u\beta\in R$ where $\alpha,\beta \in \mathbb{F}_{p^m}$.
When $\beta=0$ and $\alpha\neq 0$ arbitrary, a system description for $\lambda$-constacyclic codes
over $R$ of length $p^sn$ has been completed for any positive integer $n$ satisfying ${\rm gcd}(p,n)=1$ by
Yonglin Cao, Yuan Cao, Jian Gao and Fang-Wei Fu (FFA-15-169: Constacyclic codes
of length $p^sn$ over $\mathbb{F}_{p^m}+u\mathbb{F}_{p^m}$).


\end{document}